\documentclass[a4paper,11pt,oneside,notitlepage]{book}

\usepackage{amsmath,amssymb}

\usepackage[dvips]{graphicx}
\usepackage[font=small,labelfont=bf,labelsep=period,margin=0cm,tableposition=top]{caption}
\usepackage{subfigure}

\usepackage{natbib}
\bibpunct[]{(}{)}{;}{a}{}{,}

\usepackage{setspace}
\usepackage[lmargin=4cm,rmargin=3cm,vmargin=3cm]{geometry}

\usepackage[medium]{titlesec}

\usepackage{rotating}
\usepackage{longtable}

\usepackage{array}

\usepackage{enumerate}

\begin{document}

\frontmatter


\thispagestyle{empty}

\begin{singlespace}

\begin{center}

\rule{\textwidth}{3pt}

\vspace*{\stretch{2}}

\Huge
\bf%
The Formation and Evolution of S0 Galaxies \\

\normalsize

\vspace{\stretch{1}}

\huge%
\bf%
Alejandro~P.~Garc\'{i}a~Bedregal

\normalsize


\vspace{\stretch{8}}

\includegraphics[width=0.45\textwidth]
                {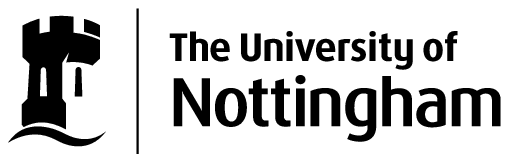}

\vspace{\stretch{0.5}}

\Large%
\bf%
Thesis submitted to the University of Nottingham\\
for the degree of Doctor of Philosophy
\bigskip\\
January 2007

\normalsize

\vspace{\stretch{1}}

\rule{\textwidth}{3pt}

\end{center}

\newpage



\thispagestyle{empty}

\noindent


\vspace*{17.5cm}

\large%

\noindent
\makebox[8em][l]{\textbf{Supervisor:}}  Dr.~Alfonso~Arag{\'o}n-Salamanca

\vspace*{2mm}

\noindent
\makebox[8em][l]{\textbf{Co-supervisor:}}  Prof.~Michael~R.~Merrifield

\vspace*{4mm}

\noindent
\makebox[8em][l]{\textbf{Examiners:}}  Prof.~Roger~Davies \quad \emph{(University of Oxford)}

\vspace*{2mm}

\noindent
\makebox[8em][l]{                   }  Dr.~Meghan~Gray \quad \emph{(University of Nottingham)}

\vspace*{12mm}

\noindent
\makebox[8em][l]{\textbf{Submitted:}}  1 December 2006

\vspace*{4mm}

\noindent
\makebox[8em][l]{\textbf{Examined:}}  16 January 2007

\vspace*{4mm}

\noindent
\makebox[8em][l]{\textbf{Final version:}}  22 January 2007

\normalsize

\end{singlespace}


\setlength{\baselineskip}{15pt}

\chapter*{Abstract}
\addcontentsline{toc}{chapter}{\numberline {}Abstract}%
This thesis studies the origin of local S0 galaxies and their possible links
to other morphological types, particularly during their evolution. To
address these issues, two different -- and complementary -- approaches have been
adopted: a detailed study of the stellar populations of S0s in
the Fornax Cluster and a study of the Tully--Fisher Relation of local S0s in
different environments.

The data utilised for the study of Fornax S0s includes new long-slit
spectroscopy for a sample of 9 S0 galaxies obtained using the FORS2
spectrograph at the 8.2m ESO VLT.  From these data, several kinematic
parameters have been extracted as a function of position along the
major axes of these galaxies. These parameters are the mean velocity, velocity
dispersion and higher-moment $h_3$ and $h_4$ coefficients.  Comparison with
published kinematics indicates that earlier data are often limited by their
lower signal-to-noise ratio and relatively poor spectral resolution.  The
greater depth and higher resolution of the new data mean that we reach
well beyond the bulges of these systems, probing their disk kinematics
in some detail for the first time.  Qualitative inspection of the
results for individual galaxies shows that some of them are not entirely
simple systems, perhaps indicating a turbulent past.  Nonetheless, circular
velocities are reliably derived for seven rotationally-supported systems of
this sample. 

The analysis of the central absorption line indices of these 9 galaxies indicates
that they correlate with central velocity dispersions ($\sigma_0$) in a way
similar to what previous studies found for ellipticals. However, the stellar
population properties of Fornax S0s indicates that the observed trends seem to
be produced by relative differences in age and $\alpha$-element abundances,
contrary to what is found in ellipticals where the overall metallicities are
the main drivers of the correlations. It was found that the observed scatter
in the line indices versus $\sigma_0$ relations can be partially explained by
the rotationally-supported nature of many of these systems. The tighter
correlations found between line indices and maximum rotational velocity support
  this statement. It was also confirmed that the dynamical mass is the driving
  physical property of all these correlations and in our Fornax S0s it has to be
  estimated assuming rotational support.

  In this thesis, a study of the local $B$- and $K_{\rm s}$-band Tully--Fisher
  Relation (TFR) in S0 galaxies is also presented.  Our new high-quality
  spectral data set from the Fornax Cluster and kinematical
  data from the literature was combined with homogeneous photometry from the RC3
  and 2MASS catalogues to construct the largest sample of S0 galaxies ever
  used in a study of the TFR.  Independent of environment, S0 galaxies are
  found to lie systematically below the TFR for nearby spirals in both the optical
  and infrared bands.  This offset can be crudely interpreted as arising from
  the luminosity evolution of spiral galaxies that have faded since ceasing
  star formation.

  However, a large scatter is also found in the S0 TFR.  Most
  of this scatter seems to be intrinsic, not due to the observational
  uncertainties.  The presence of such a large scatter means that the
  population of S0 galaxies cannot have formed exclusively by the
  above simple fading mechanism after all transforming at a single
  epoch.
  
  To better understand the complexity of the transformation mechanism,
  a search for correlations was carried out between the offset from the TFR
  and other properties of the galaxies such as their structural
  properties, central velocity dispersions and ages (as
  estimated from absorption line indices). For the Fornax Cluster data, the
  offset from the TFR correlates with the estimated age of the stars
  in the centre of individual galaxies, in the sense and of the magnitude
  expected if S0 galaxies had passively faded since being converted
  from spirals.  This correlation could imply that part of the
  scatter in the S0 TFR arises from the different times at which galaxies began
  their transformation.




\chapter*{Published work}
\addcontentsline{toc}{chapter}{\numberline {}Published work}

Much of the work in this thesis has been previously presented in two papers:
\begin{enumerate}[1.]
\item Bedregal et al.\ (2006a), ``S0 galaxies in Fornax: data and
  kinematics''.

\item Bedregal et al.\ (2006b), ``The Tully-Fisher relation for S0 galaxies''.

\end{enumerate}

The rest will be presented in:
\begin{enumerate}[3.]
\item Bedregal et al.\ (2007), ``S0 galaxies in Fornax: Central Stellar populations'' (submitted to MNRAS).

\end{enumerate}
Bedregal et al.\ (2006a) contains much of the work detailed in Chapter~2.
Bedregal et al.\ (2006b) describes the work presented in Chapter~4.
The contents of Chapter~3 will be presented in
Bedregal et al.\ (2007).
The work presented in this thesis was performed by the
author, with advice from the paper coauthors listed above.  Where the material
presented is taken from literature, this is mentioned explicitly in the relevant chapter.
\\ \\
Finally, other work performed during the PhD which \emph{is not} included in
this thesis has being publish in two other papers:
\begin{enumerate}[4.]

\item Arag{\'o}n-Salamanca et al.\ (2006), ``Measuring the fading of S0 galaxies using globular clusters''.
\end{enumerate}
\begin{enumerate}[5.]

\item Barr et al.\ (2007), ``The Formation of S0 galaxies: evidence from globular clusters'' (submitted to A\&A).  

\end{enumerate}


\mainmatter

\pagestyle{headings}







\begin{thebibliography}{}

\bibitem{A06}{Arag{\'o}n-Salamanca A., Bedregal A.G. \& Merrifield M.R.},
  2006, A\&A, 458, 101
\bibitem{B06a}{Bedregal A.G., Arag\'on-Salamanca A., Merrifield M.R.,
  Milvang-Jensen B.}, 2006a, MNRAS, 371, 1912
\bibitem{B06b}{Bedregal A.G., Arag{\'o}n-Salamanca A. \& Merrifield M.R.},
  2006b, MNRAS, 373, 1125

\end{thebibliography}


\backmatter

\end{document}